\documentclass[12pt]{article}

\begin{document}

\baselineskip 0.75cm
\topmargin -0.6in
\oddsidemargin -0.1in

\let\ni=\noindent

\renewcommand{\thefootnote}{\fnsymbol{footnote}}

\newcommand{\CKM}{Cabibbo--Kobayashi--Maskawa }

\newcommand{\SM}{Standard Model }

\pagestyle {plain}

\setcounter{page}{1}

\pagestyle{empty}

~~~

\begin{flushright}
{\bf IFT-01/01\\
\bf hep-ph/0101014}
\end{flushright}

\vspace{2.cm}

{\large\centerline{\bf Possible LSND effect as a small perturbation}}
{\large\centerline{\bf of the bimaximal texture for three active neutrinos}}


\vspace{0.3cm}

{\centerline {\sc Wojciech Kr\'{o}likowski}}

\vspace{0.3cm}

{\centerline {\it Institute of Theoretical Physics, Warsaw University}}

{\centerline {\it Ho\.{z}a 69,~~PL--00--681 Warszawa, ~Poland}}

\vspace{1.8cm}

{\centerline{\bf Abstract}}

\vspace{0.3cm}
A particular form of mixing matrix for three active and one sterile neutrinos
is proposed. Its $3\times 3 $ part describing three active neutrinos arises 
from the popular bimaximal mixing matrix  that works satisfactorily in solar 
and atmospheric experiments if the LSND effect is ignored. Then, the sterile 
neutrino, effective in the fourth row and fourth column of the proposed mixing 
matrix, is responsible for the possible LSND effect by inducing one extra 
neutrino mass state to exist actively. The LSND effect, if it exists, turns out
to reveal its {\it perturbative} nature related to small mixing of three active
neutrinos with their sterile partner.

\vspace{0.3cm}

\ni PACS numbers: 12.15.Ff, 14.60.Pq, 12.15.Hh.

\vspace{0.6cm}

\ni January 2001

\vfill\eject

~~~
\pagestyle {plain}

\setcounter{page}{1}

\vspace{0.2cm}

Recent experimental results for atmospheric $\nu_\mu $'s as well as solar 
$\nu_e $'s favour excluding the hypothetical sterile neutrinos from neutrino 
oscillations [1]. However, the problem of the third neutrino mass-square 
difference, related to the possible LSND effect for accelerator $\nu_\mu $'s,  
still exists [2], stimulating a further discussion about mixing of three active
neutrinos  with their sterile counterparts. In the present note we contribute 
to this discussion by constructing a particular $4\times 4$ texture of three 
active and  one sterile neutrinos, $ \nu_e\,, \,\nu_\mu\,, \,\nu_\tau $ and 
$\nu_s $, whose $3\times 3$ part describing three active neutrinos arises from 
the popular bimaximal texture that works in a satisfactory way in solar and 
atmospheric experiments if the LSND effect is ignored [3].
 
In this popular $3\times 3$ texture the mixing matrix has the form

\begin{equation} 
U^{(3)} =   \left( \begin{array}{ccc} c_{12} & s_{12} & 0 \\
- {s_{12}}/{\sqrt{2}} & \;\,{c_{12}}/{\sqrt{2}} & 1/\sqrt{2} \\ 
\;\,{s_{12}}/{\sqrt{2}} & -{c_{12}}/{\sqrt{2}} & 1/\sqrt{2}\end{array}\right) 
\;,
\end{equation} 

\ni where $ c_{12} \simeq 1/\sqrt{2} \simeq s_{12}$. Such a form corresponds to
$c_{23} = 1/\sqrt{2} = s_{23}$ and $ s_{13} = 0$ in the notation usual for a 
general \CKM$\!\!$--type matrix [4] (if the LSND effect is ignored, the upper 
limit $|s_{13}| < 0.1 $ follows from the negative result of Chooz reactor 
experiment [5]). Then, in the $4\times 4$ texture we proposed the following 
mixing matrix:

\begin{equation} 
U=\left(\begin{array}{cccc} c_{14} c_{12} & s_{12} & 0 & \;\,s_{14}c_{12}\\
-c_{14}s_{12}/{\sqrt{2}}&\;\,c_{12}/{\sqrt2}&1/\sqrt{2}&-s_{14}s_{12}/\sqrt{2}\\
\;\,c_{14} s_{12} /{\sqrt{2}} & -c_{12}/{\sqrt{2}} & 1/\sqrt2 & \;\,s_{14} 
s_{12}/\sqrt{2} \\ -s_{14} & 0 & 0 & c_{14} \end{array} \right) 
\end{equation} 

\vspace{0.1cm}

\ni with $ c_{ij} = \cos \theta_{ij}$ and $ s_{ij} = \sin \theta_{ij}$. Here, 
$U = \left(U_{\alpha i } \right)\;,\;\alpha = e\,,\,\mu\,,\,\tau\,,\,s$ and 
$i = 1,2,3,4$, while the unitary transformation describing the mixing of four 
neutrinos $ \nu_e\,, \,\nu_\mu\,, \,\nu_\tau $ and $ \nu_s $ is inverse to the 
form 

\begin{equation}
\nu_\alpha = \sum_i U_{\alpha i} \nu_i \;,
\end{equation}

\ni where $ \nu_1\,, \,\nu_2\,, \,\nu_3 $ and $ \nu_4 $ denote the massive 
neutrinos carrying the masses $ m_1\,, \,m_2\,, \,m_3$ and $ m_4 $. Of course, 
$U^\dagger = U^{-1}$ and $U^* = U$. Note that

\begin{equation} 
 U = \left( \begin{array}{cccc} \, & \, & \, & 0 \\ \, & {U^{(3)}} & \, & 0 
\\ \, & \, & \, & 0  \\ 0 & 0 & 0 & 1 \end{array} \right) + O(s_{14}) 
\end{equation} 

\vspace{0.2cm}

\ni in the limiting case of $ |s_{14}| \ll |c_{14}|$. Since in the limit of 
$ s_{14} = 0 $ there is no LSND effect, Eq. (4) suggests that this possible 
effect has a {\it perturbative} character, consistent with its small estimated 
amplitude $\sin^2 2\theta_{\rm LSND} \sim 10^{-2}$.

It is interesting to consider a $6\times 6$ texture of three active and three 
sterile neutrinos which may be the active and conventional--sterile Majorana 
neutrinos, $\nu^{(a)}_\alpha\equiv\nu_{\alpha L}+\left(\nu_{\alpha L}\right)^c$ 
and $\nu^{(s)}_\alpha\equiv\nu_{\alpha R}+\left(\nu_{\alpha R}\right)^c ,\, 
\alpha = e\,,\,\mu\,,\,\tau$. Define in this texture the following mixing 
matrix  

\vspace{0.1cm}

\begin{equation}
U^{(6)} = \left(\begin{array}{cc} U^{(3)} & 0^{(3)} \\ 0^{(3)} & 1^{(3)}   
\end{array} \right) \left( \begin{array}{rc} C & S \\ -S & C \end{array} 
\right)  \;,\; C =  \left( \begin{array}{ccc} c_{14} & 0 & 0 \\ 0 & c_{25} 
& 0 \\ 0 & 0 & c_{36} \end{array} \right) \;,\;  S =  
\left( \begin{array}{ccc} s_{14} & 0 & 0 \\ 0 & s_{25} & 0 \\ 0 & 0 & s_{36} 
\end{array} \right) \;.
\end{equation} 

\vspace{0.1cm}

\ni Then, it is easy to discover that

\vspace{0.1cm}

\begin{equation} 
U^{(6)} = \left( \begin{array}{cccccc} \, & \, & \, & \, & 0 & 0 \\ 
\, & \, & \, & \, & 0 & 0 \\ \, & \, & U & \, & 0 & 0 \\ 
\, & \, & \, & \, & 0  & 0  \\ 0 & 0 & 0& 0 & 1& 0  \\ 
0 & 0  & 0  & 0  & 0  & 1 \end{array} \right) + O(s_{25}) + O(s_{36}) 
\end{equation} 

\vspace{0.2cm}

\ni in the limiting case of $|s_{25}|\ll |c_{25}|$ and $|s_{36}| \ll |c_{36}|$.
In this case, two sterile neutrinos $\nu^{(s)}_\mu $ and $\nu^{(s)}_\tau $ 
become decoupled from three active neutrinos 
$\nu^{(a)}_e \,,\, \nu^{(a)}_\mu \,,\, \nu^{(a)}_\tau $ and from one sterile 
neutrino $\nu^{(s)}_e $, if our $ 6\times 6$ texture is realized indeed with 
the use of three active and three conventional--sterile Majorana neutrinos. 
Then, four neutrinos $\nu^{(a)}_e \,,\, \nu^{(a)}_\mu \,,\, \nu^{(a)}_\tau $ 
and $\nu^{(s)}_e $ mix through the matrix inverse to $U$ given in Eq. (2).

In the representation, where the mass matrix of three charged leptons 
$e^- \,,\, \mu^-  \,,\, \tau^- $ is diagonal, the $4\times 4 $ neutrino 
mixing matrix $U$ is at the same time the diagonalizing matrix for the 
$4\times 4$ neutrino mass matrix $M = \left(M_{\alpha \beta} \right)$:

\begin{equation} 
U^\dagger M U = {\rm diag}(m_1\,,\,m_2\,,\,m_3\,,\,m_4)\;.
\end{equation} 

\ni Here, by definition $ m_1 \leq m_2 \leq m_3$ and either $ m_3 \leq m_4$ 
or $ m_4 \leq m_1$. Then, evidently 
$ M_{\alpha \beta} = \sum_i U_{\alpha i} m_i U^*_{\beta i}$, and hence with 
the use of proposal (2) we obtain

\vspace{0.1cm}

\begin{eqnarray} 
M_{e e} & = & c^2_{12}\left( c^2_{14} m_1 + s^2_{14} m_4\right) + s^2_{12} m_2 
\;, \nonumber \\
M_{e \mu} & = & - M_{e \tau} = - \frac{1}{\sqrt{2}}\, c_{12}\, s_{12} 
\left( c^2_{14} m_1 + s^2_{14} m_4 - m_2 \right)\;, \nonumber \\ 
M_{e s} & = & - c_{12}\, c_{14}\, s_{14} \left( m_1- m_4 \right) \;, 
\nonumber \\ 
M_{\mu \mu} & = & M_{\tau \tau} = \frac{1}{2} \left[s^2_{12} 
\left( c^2_{14} m_1 + s^2_{14} m_4\right) + c^2_{12} m_2 + m_3 \right] \;, 
\nonumber \\ 
M_{\mu \tau} & = & - \frac{1}{2} \left[ s^2_{12}
\left( c^2_{14} m_1 + s^2_{14} m_4\right) + c^2_{12} m_2 - m_3 \right] \;, 
\nonumber \\ 
 M_{\mu s} & = & - M_{\tau s} = \frac{1}{\sqrt{2}} s_{12}\, c_{14}\, s_{14} 
\left( m_1 - m_4 \right) \;, \nonumber \\  
M_{s s} & = & s^2_{14} m_1 + c^2_{14} m_4 \;,
\end{eqnarray}

\vspace{0.1cm}

\ni where $ c_{12} \simeq 1/\sqrt{2} \simeq s_{12}\;\; i.\,e., \;\; 
\theta_{12} \simeq \pi /4$. Of course, $M^\dagger = M $ and $ M^* = M $.

Due to mixing of four neutrino fields described by Eq. (3), neutrino 
states mix according to the relation

\vspace{0.1cm}

\begin{equation} 
|\nu_\alpha \rangle = \sum_i U^*_{\alpha i} |\nu_i \rangle \;.
\end{equation}

\vspace{0.1cm}

\ni This implies the following familiar formulae for probabilities of neutrino 
oscillations $ \nu_\alpha \rightarrow \nu_\beta $ on the energy shell:

\begin{equation} 
P(\nu_\alpha \rightarrow \nu_\beta) = 
|\langle \nu_\beta| e^{i PL} |\nu_\alpha  \rangle |^2 = 
\delta _{\beta \alpha} - 4\sum_{j>i} U^*_{\beta j} 
U_{\beta i} U_{\alpha j} U^*_{\alpha i} \sin^2 x_{ji} \;,
\end{equation}

\ni being valid if the quartic product 
$ U^*_{\beta j} U_{\beta i} U_{\alpha j} U^*_{\alpha i}$ is real, what is 
certainly true when the tiny CP violation is ignored. Here,


\begin{equation} 
x_{ji} = 1.27 \frac{\Delta m^2_{ji} L}{E} \;\;,\;\;  
\Delta m^2_{ji} = m^2_j - m^2_i
\end{equation}

\vspace{0,1cm}

\ni with $\Delta m^2_{ji}$, $L$ and $E$ measured in eV$^2$, km and GeV, 
respectively ($L$ and $E$ denote the experimental baseline and neutrino 
energy, while $ p_i = \sqrt{E^2 - m_i^2} \simeq E -m^2_i/2E $ are eigenvalues 
of the neutrino momentum $P$).

With the use of proposal (2) for the $ 4\times 4$ neutrino mixing matrix the 
oscillation formulae (10) lead to the probabilities

\vspace{0.2cm}

\begin{eqnarray}
P(\nu_e\, \rightarrow \nu_e)\! & \!\simeq\! & \!1 - (2c_{12} s_{12})^2 
c^2_{14} \sin^2 x_{21} - 4(1 - c^2_{12} s^2_{14})c^2_{12} s^2_{14} 
\sin^2 x _{41} \;, \nonumber \\
P( \nu_\mu\! \rightarrow \nu_\mu)\! & \!=\! & \!P( \nu_\tau \rightarrow 
\nu_\tau)  \simeq 1 - (c_{12}s_{12})^2 c^2_{14} \sin^2 x _{21}\nonumber \\ 
& & - (1 -s^2_{12} s^2_{14}) \left( \sin^2 x _{32} +s^2_{12} s_{14}^2 
\sin^2 x _{41}\right) - s^2_{12} s_{14}^2 \sin^2 x _{43} \,, \nonumber \\ 
P(\nu_\mu \rightarrow \nu_e)\! & \!=\! & \!P(\nu_\tau \rightarrow \nu_e) 
\simeq 2(c_{12} s_{12})^2 \left( c^2_{14} \sin^2 x_{21} + s^4_{14}  
\sin^2 x _{41}\right) \;, \nonumber \\
P(\nu_\mu \rightarrow \nu_\tau)\! & \!\simeq\! & \!- (c_{12}s_{12})^2 
c^2_{14} \sin^2 x_{21} + (1 -s^2_{12} s^2_{14}) \left( \sin^2 
x _{32} - s^2_{12} s_{14}^2 \sin^2 x _{41}\right) \nonumber \\ 
& &  + s^2_{12} s_{14}^2 \sin^2 x _{43} 
\end{eqnarray}

\ni in the approximation where $m^2_1 \simeq m^2_2$ (and both are much 
different from $m^2_3$ and  $m^2_4$), and also to the probabilities involving 
the sterile neutrino $ \nu_s$

\begin{eqnarray}
P(\nu_\mu \rightarrow \nu_s) & = & P(\nu_\tau \rightarrow \nu_s) = 2 s^2_{12} 
(c_{14} s_{14})^2 \sin^2 \!x _{41} \;, \nonumber \\
P(\nu_e \rightarrow \nu_s) & = & 4 c^2_{12} (c_{14} s_{14})^2 \sin^2x _{41} 
\;, \nonumber \\ 
P(\nu_s \rightarrow \nu_s) & = & 1 - 4 (c_{14} s_{14})^2 \sin^2 \!x _{41}
\end{eqnarray}

\ni where only $ m^2_1 $ and $ m^2_4 $ participate.

If $|\Delta m^2_{21}| \ll |\Delta m^2_{41}|$ ({\it i.e.}, 
$|x_{21}| \ll |x_{41}|$) and

\begin{equation} 
|\Delta m^2_{21}| = \Delta m^2_{\rm sol} \sim 
(10^{-5}\;\;{\rm or}\;\;10^{-7} \;\; {\rm or}\;\; 10^{-10})\;{\rm eV}^2  
\end{equation} 

\ni for LMA or LOW or VAC solution, respectively [1,6], then under the 
conditions of solar experiments the first Eq. (12) with 
$ c_{12} \simeq 1/\sqrt{2} \simeq s_{12}$ gives 

\begin{equation} 
P(\nu_e \rightarrow \nu_e)_{\rm sol} \simeq 1 -  c^2_{14} 
\sin^2 (x _{21})_{\rm sol} - \frac{(1+c_{14}^2) s_{14}^2}{2}\;,\; c^2_{14} = 
\sin^ 2 2\theta_{\rm sol} \sim 0.8\;{\rm or}\; 0.9\;{\rm or}\; 0.7  \;.  
\end{equation} 

If $|\Delta m^2_{21}| \ll |\Delta m^2_{32}| \ll 
|\Delta m^2_{41}|\,,\, |\Delta m^2_{43}|$ 
({\it i.e.}, $|x_{21}| \ll |x_{32}| \ll |x_{41}| \,,\, |x_{43}|$) and 

\begin{equation} 
|\Delta m^2_{32}| = \Delta m^2_{\rm atm} \sim 3 \times10^{-3}\;{\rm eV}^2 \;, 
\end{equation} 

\ni then for atmospheric experiments the second Eq. (12) with $ c_{12} 
\simeq 1/\sqrt{2} \simeq s_{12}$  leads to

\begin{equation} 
P( \nu_\mu \rightarrow \nu_\mu)_{\rm atm}  \simeq  1 - \frac{1 + c^2_{14}}{2} 
\sin^2 (x _{32})_{\rm atm} - \frac{(3+c^2_{14}) s_{14}^2}{8}
\;\;,\;\;\frac{1 + c^2_{14}}{2}  = \sin^2 2\theta_{\rm atm} \sim 1\;.  
\end{equation} 

Eventually, if $|\Delta m^2_{21}| \ll |\Delta m^2_{41}|$ and 

\begin{equation} 
|\Delta m^2_{41}| = \Delta m^2_{\rm LSND} \sim 1 \;{\rm eV}^2 \;\;(e.g.) \;,
\end{equation} 

\ni then in the LSND experiment the third Eq. (12) with $ c_{12} 
\simeq 1/\sqrt{2} \simeq s_{12}$  implies

\begin{equation} 
P( \nu_\mu \rightarrow \nu_e)_{\rm LSND}  \simeq 
\frac{s^4_{14}}{2}\sin^2 (x _{41})_{\rm LSND}\;\;,\;\;\frac{s^4_{14}}{2} = 
\sin^2 2\theta_{\rm LSND} \sim 10^{-2}\;\;(e.g.).  
\end{equation} 

\ni Thus,

\begin{equation} 
s_{14}^2 \sim 0.14 \,,\, c_{14}^2 \sim 0.86 \,,\, \frac{1 + c^2_{14}}{2}  
\sim 0.93 \,,\, \frac{(1+c_{14}^2) s_{14}^2}{2} \sim 0.13\,,\,
\frac{(3+c_{14}^2) s_{14}^2}{8} \sim 0.068\,,
\end{equation} 

\ni if the LSND effect really exists and develops the amplitude 
$s^4_{14}/2 \sim 10^{-2}$. Through Eq. (19) the LSND effect (if it exists) 
reveals its {\it perturbative} nature related to the small constant 
$ s^4_{14}/2 \sim 10^{-2}$ that measures coupling of $\nu_1$ with $\nu_4$ in 
the process of $\nu_\mu \rightarrow \nu_e $ oscillations at LSND.

Concluding, we can say that Eqs. (15), (17) and (19) are not inconsistent 
with solar, atmospheric and LSND experiments, respectively. Note that in 
Eqs. (15) and (17) there are constant terms that modify moderately the usual 
two--flavor formulae. Any LSND--type accelerator experiment, in contrast to 
the solar and atmospheric projects, investigates a small 
$\nu_\mu \rightarrow \nu_e$ oscillation effect caused possibly by the sterile 
neutrino. So, this effect (if it exists) plays the role of a small 
{\it perturbation} of the basic bimaximal texture for three active neutrinos.

The final equations (15), (17) and (19) follow from the first three 
oscillation formulae (12), if either

\begin{equation} 
m^2_1 \simeq m^2_2 \ll m^2_3 \ll m^2_4 
\end{equation} 

\ni with

\begin{equation} 
m^2_3 \ll 1\;\;{\rm eV}^2\;\;,\;\; m^2_4 \sim 1\;\;{\rm eV}^2\;\;,\;\; 
\Delta m^2_{21} \sim (10^{-5} -10^{-10})\;{\rm eV}^2 \ll \Delta m^2_{32} 
\sim 10^{-3} \;{\rm eV}^2 
\end{equation} 

\ni or

\begin{equation} 
m^2_1 \simeq m^2_2 \simeq m^2_3 \gg m^2_4 
\end{equation} 

\ni with

\begin{equation} 
m^2_1 \sim 1\;\;{\rm eV}^2\;\;,\;\; m^2_4 \ll 1\;\;{\rm eV}^2\;\;,\;\; 
\Delta m^2_{21} \sim (10^{-5} -10^{-10})\;{\rm eV}^2 \ll \Delta m^2_{32} 
\sim 10^{-3} \;{\rm eV}^2 \;.
\end{equation} 

\ni Indeed, when either $ m^2_1 \simeq m^2_2 \ll m^2_3 \ll m^2_4 
\sim 1\;{\rm eV}^2$ or $m^2_4 \ll m^2_1 \simeq m^2_2 \simeq m^2_3 
\sim 1\;{\rm eV}^2$, we may obtain $\Delta m^2_{21} \ll \Delta m^2_{32} 
\ll |\Delta m^2_{41}| \sim 1\;{\rm eV}^2 $. The second case of $ m^2_4 
\ll m^2_1 \sim 1 \;{\rm eV}^2 $, where the neutrino mass state $i = 4$ induced 
by the sterile neutrino $\nu_s$ gets a vanishing mass, seems to be more natural
 than the first case of $ m_3^2 \ll m^2_4 \sim 1 \;{\rm eV}^2 $, where such a 
state gains a considerable amount of mass "for nothing". This is so, unless 
one believes in the liberal maxim "whatever is not forbidden is allowed": the 
Majorana righthanded mass is not forbidden by the electroweak SU(2)$\times$U(1) symmetry, in contrast to Majorana lefthanded and Dirac masses requiring this 
symmetry to be broken (for the active Majorana neutrinos), say, by a Higgs 
mechanism that becomes then the origin of these masses. Here, the active 
Majorana neutrinos are $\nu^{(a)}_\alpha \equiv \nu_{\alpha L} + 
(\nu_{\alpha L})^c\;,\; \alpha = e\,,\,\mu\,,\,\tau $, while the sterile 
Majorana neutrino is $\nu_s \equiv  \nu_{s R} + (\nu_{s R})^c$ with 
$\nu_{s L} =  (\nu_{s R})^c = (\nu_s^c)_L$ (implying effectively the Dirac 
$1\times 3 $ mass matrix and the Dirac transposed $3\times 1$ mass matrix as 
well as the Majorana righthanded trivial $1\times 1$ mass matrix). Possibly 
$\nu_s = \nu_e^{(s)}$ [{\it cf.} the comment to Eq. (6)]; then 
$\nu_{s R} = \nu_{e R}$.

In the approximation used before to derive Eqs. (15), (17) and (19) there are 
true also the relations

\begin{eqnarray} 
P(\nu_e\, \rightarrow\, \nu_e)_{\rm sol} & \simeq & \!1 - \! P( \nu_e 
\rightarrow \nu_\mu)_{\rm sol}\! - \! 
P( \nu_e \rightarrow \nu_\tau)_{\rm sol}\! - \!(c_{14} s_{14})^2 \;, 
\; (c_{14} s_{14})^2 \sim 0.12 \;, \nonumber \\
P( \nu_\mu\! \rightarrow \nu_\mu)_{\rm atm} & \simeq & 1 - 
P( \nu_\mu \rightarrow \nu_\tau)_{\rm atm} -  
\frac{(3 + c^2_{14})s^2_{14}}{8}\;\; , \;\; \frac{(3 + c^2_{14}) s^2_{14}}{8} 
\sim 0.068 \;,  
\end{eqnarray} 

\ni as well as

\begin{equation} 
P(\nu_\mu \rightarrow \nu_e)_{\rm LSND} \simeq \frac{1}{2}
\left( \frac{s_{14}}{c_{14}} \right)^2 P(\nu_\mu \rightarrow \nu_s)_{\rm LSND}
\;\;,\; \frac{1}{2}\left( \frac{s_{14}}{c_{14}}\right)^2 \sim 0.082 \;\;.\;\;
\end{equation}

\ni The second relation (25) demonstrates a leading role of the appearance 
mode $\nu_\mu \rightarrow \nu_\tau $ in the disappearance process of 
atmospheric $\nu_\mu$'s, while the relation (26) indicates a direct interplay 
of the appearance modes $\nu_\mu \rightarrow \nu_e $ and 
$\nu_\mu \rightarrow \nu_s $. In the case of the first relation (25), both 
appearance modes $\nu_e \rightarrow \nu_\mu $ and $\nu_e \rightarrow \nu_\tau$ 
contribute equally to the disappearance process of solar $\nu_e$'s, and the 
role of the appearance mode $\nu_e \rightarrow \nu_s$ (responsible for the 
constant term) is also considerable.

Finally, for the Chooz experiment [5], where it happens that 
$(x_{ji})_{\rm Chooz} \simeq (x_{ji})_{\rm atm}$, the first Eq. (12) predicts 

\begin{equation} 
P(\bar{\nu}_e \rightarrow\, \bar{\nu}_e)_{\rm Chooz}  \simeq  P( \bar{\nu}_e 
\rightarrow \bar{\nu}_e)_{\rm atm} \simeq 1 -  
\frac{(1 + c^2_{14})s^2_{14}}{2} \;, \;  \frac{(1 + c^2_{14})s^2_{14}}{2} 
\sim 0.13 \;,
\end{equation} 

\ni if there is the LSND effect with the amplitude $s^4_{14}/2 \sim 10^{-2}$ 
as written in Eq. (19). Here, 
$ (1 + c^2_{14})s^2_{14}\sin^2(x_{41})_{\rm Chooz} 
\simeq  (1 + c^2_{14})s^2_{14}/2$. In terms of the usual two--flavor formula, 
the negative result of Chooz reactor experiment excludes the disappearance 
process of reactor $ \bar{\nu}_e$'s for moving $\sin^2 2\theta_{\rm Chooz} 
\stackrel{>}{\sim} 0.1 $, when the range of moving $\Delta m^2_{\rm Chooz} 
\stackrel{>}{\sim} 3\times 10^{-3}\;{\rm eV}^2 $ is considered. In our case 
$\sin^2 2\theta_{\rm Chooz} \sim  (1 + c^2_{14})s^2_{14}/2$ for 
$\sin^2 x_{\rm Chooz} \sim 1$. Thus, the Chooz effect for reactor 
$\bar{\nu}_e$'s may appear at the edge (if the LSND effect really exists). 

From the neutrinoless double $\beta$ decay, not observed so far,  the 
experimental bound $ M_{ee} \equiv \sum_i U^2_{e i} m_i < 
[0.4 (0.2)\;-\; 1.0 (0.6)]$ eV follows [7]. On the other hand, with 
$c_{12} \simeq 1/\sqrt{2} \simeq s_{12}$ and the values (20) the first 
Eq. (8) gives

\begin{equation} 
M_{ee} \sim \frac{1}{2}(0.86 m_1 + 0.14 m_4 + m_2)\,,
\end{equation}

\ni what in the case of Eq. (21) with $ m^2_4 \sim 1\;\,{\rm eV}^2$ or Eq. 
(23) with  $ m^2_1 \sim 1\;\,{\rm eV}^2$ leads to the estimate $M_{ee} 
\sim 0.07 m_4 \sim 0.07$ eV or $M_{ee} \sim 0.9 m_1 \sim 0.9$ eV, respectively.
 Of course, the value $(m^2_4$ or $ m^2_1) \simeq \Delta m^2_{\rm LSND} 
\sim 1\;{\rm eV}^2$ in Eq. (18) is only an example, and may turn out to 
be smaller.

\vfill\eject

~~~~
\vspace{0.5cm}

{\centerline{\bf References}}

\vspace{0.5cm}

{\everypar={\hangindent=0.5truecm}
\parindent=0pt\frenchspacing

{\everypar={\hangindent=0.5truecm}
\parindent=0pt\frenchspacing

1.~{\it Cf. e.g.} E. Kearns, Plenary talk at {\it ICHEP 2000} at 
Osaka; C.~Gonzales--Garcia, Talk at {\it ICHEP 2000} at Osaka.

\vspace{0.2cm}

2.~G. Mills, Talk at {\it Neutrino 2000}; R.L.~Imlay, Talk at 
{\it ICHEP 2000} at Osaka; and references therein.

\vspace{0.2cm}

3.~{\it Cf. e.g.} F. Feruglio, {\it Acta Phys. Pol.} {\bf B 31}, 1221 
(2000); J.~Ellis, Summary of {\it Neutrino 2000}, hep-ph/0008334; and 
references therein.

\vspace{0.2cm}

4.~W. Kr\'{o}likowski, hep--ph/0007255; hep--ph/0010331.

\vspace{0.2cm}

5.~M. Appolonio {\it et al.}, {\it Phys. Lett.} {\bf B 420}, 397 (1998); 
{\bf B 466}, 415 (1999).

\vspace{0.2cm}

6.~M.V.~Garzelli and C.~Giunti, hep--ph/0012247.

\vspace{0.2cm}

7.~L.~Baudis {\it et al.}, {\it Phys. Rev. Lett.} {\bf 83}, 41 (1999); 
{\it cf.} also Review of Particle Physics, {\it Eur. Phys. J.} {\bf C 15}, 1 
(2000), p. 363.

\vfill\eject

\end{document}